# A genome wide dosage suppressor network reveals genetic robustness and a novel mechanism for Huntington's disease

Biranchi Patra[1]*, Yoshiko Kon[2]*, Gitanjali Yadav[1,3], Anthony W. Sevold[1], Jesse P. Frumkin[1], Ravishankar R. Vallabhajosyula[1,¶], Arend Hintze[1,□], Bjørn Østman[1,□], Jory Schossau[1,$], Ashish Bhan[1], Bruz Marzolf[4], Jenna K. Tamashiro[1], Amardeep Kaur[4], Nitin S. Baliga[4], Elizabeth J. Grayhack[2], Christoph Adami[1,□], David J. Galas[4,@], Alpan Raval[1,5,#], Eric M. Phizicky[2], and Animesh Ray[1].

[1]Keck Graduate Institute, 535 Watson Drive, Claremont, CA 91711, USA; [2]Department of Biochemistry, University of Rochester School of Medicine, Rochester, NY 14627, USA; [3]National Institute of Plant Genome Research (NIPGR), Aruna Asaf Ali Marg, New Delhi, India; [4]Institute for Systems Biology, 1441 N 34th St, Seattle, WA 98103 Seattle, USA; [5]Institute of Mathematical Sciences, Claremont Graduate University, Claremont, CA 91711, USA.

[¶]Present address: CFD Research Corporation, Huntsville, Alabama, USA. [$]Present address: Pacific Northwest Diabetes Research Institute, Seattle, WA 98122. [#]Present address: D E Shaw Research, Hyderabad 500034, India. [$]Present address: Computer Science and Engineering, Michigan State University East Lansing, MI 48823, USA. [□]Present address: Microbiology and Molecular Genetics, and BEACON Center for the Study of Evolution in Action, Michigan State University, East Lansing, MI 48823, USA. [@]Present address: Pacific Northwest Diabetes Research Institute, 720 Broadway Seattle, WA 98122, USA.

*These authors made equal contributions

**Contact:** Dr. Animesh Ray    Email: aray@kgi.edu; Phone: (909) 607 9729

**Running head:** Dosage suppressor network

**Abstract**

**Mutational robustness is the extent to which an organism has evolved to withstand the effects of deleterious mutations. We explored the extent of mutational robustness in the budding yeast by genome wide dosage suppressor analysis of 53 conditional lethal mutations in cell division cycle and RNA synthesis related genes, revealing 660 suppressor interactions of which 642 are novel. This collection has several distinctive features, including high co-occurrence of mutant-suppressor pairs within protein modules, highly correlated functions between the pairs, and higher diversity of functions among the co-suppressors than previously observed. Dosage suppression of essential genes encoding RNA polymerase subunits and chromosome cohesion complex suggest a surprising degree of functional plasticity of macromolecular complexes and the existence of degenerate pathways for circumventing potentially lethal mutations. The utility of dosage-suppressor networks is illustrated by the discovery of a novel connection between chromosome cohesion-condensation pathways involving homologous recombination, and Huntington's disease.**

Biological interaction networks are robust to perturbation[1,2,3,4,5] because of several features, including power-law network topology, redundancy, modularity, and their dynamic properties[2,6,7,8,9,10,11,12,13,14,15,16,17]. Although modularity is a common feature of interaction networks[12,18,19,20,21,22], the contribution of modularity to genetic robustness is difficult to determine. Recent studies have revealed dynamic interaction among apparently unrelated gene modules in response to genotoxic stress, suggesting the existence of highly reconfigurable networks of gene and protein modules as well as of unexpectedly plastic macromolecular complexes[19]. Rewiring of signaling and/or metabolic networks have been observed in cancer cells that evolved chemotherapy resistance[23,24]. It has been speculated that genetic and epigenetic changes could accomplish network rewiring[25,26]. Complex genomic changes reminiscent of functional rewiring are associated with rapid evolutionary adaptation in response to mutation in an essential gene[27].

The suppression of essential gene mutations has been classically employed to investigate gene function, and suppressors provide clues to mechanisms of evolution[28]. More recently, genome wide dosage suppressor discovery has contributed novel insights into biological processes[29,30]. By seeking dosage suppressors of 53 mutant genes, mostly encoding cell cycle/DNA replication related genes or RNA polymerase/RNA modification related genes, we have uncovered 660 pairs of "dosage-suppressor" interactions, which supplement 1,626 dosage-suppressor pairs reported previously in the literature[31] and 254 interactions that were discovered in a recent high throughput experiment[29]. The set of interaction pairs discovered in this study overlaps significantly with known protein-protein and genetic interaction pairs, suggesting related molecular mechanisms that connect the suppressed mutant gene product with the suppressor gene's product. The mutant-suppressor pairs tend to belong to the same protein-protein interaction modules as determined by several independent criteria. We confirm all previously noted mechanisms of suppression[29], including stabilization of heat sensitive macromolecular complexes by over-producing a physically interacting protein partner (Fig. 1a), stabilization of unfolded heat sensitive proteins by chaperones, and modulation of expression of other genes participating in the same biological process as affected by the mutant gene. Here we present evidence of at least one additional mechanism of network robustness, namely, the existence of degenerate pathways. The utility of dosage-suppression network is illustrated by the serendipitous discovery of a novel connection between chromosome cohesion-condensation pathways and the toxic effects of the Huntington's disease protein.

## Results

**A genome-wide screen for dosage suppressors**

We transformed 108 isogenic yeast strains each containing a temperature sensitive (*ts*) point mutation (85 mutants) or a *ts* deletion mutation (23 mutants)[32] with pools of the entire MORF (Movable Open Reading Frame) library plasmids (Materials and Methods, Supplementary Table S1). The use of this library permits interrogating every conditional mutant with virtually every yeast ORF, under conditions where each ORF is expressed in the presence of galactose or glucose (Fig. 1b-d). We transformed each mutant strain with high copy 2μ based plasmids expressing open reading frames (ORFs) under pGAL promoter control[33], selected transformants at the permissive temperature ($25^0$C), then tested for the ability of the recombinant plasmids to rescue growth defect above the specific restrictive temperature for the corresponding mutant strain (see Methods). Each candidate suppressor ORF was individually confirmed by multiple retransformation experiments, and a subset by sequencing. In this manner we obtained 660 confirmed dosage suppressor interactions for 53 of the 108 *ts* mutants we tested (Fig. 1c, Supplementary Table S2), involving 517 suppressor genes, all of which were individually confirmed through retransformation and repeats of the assays, with all suppressors of a third of the mutant strains collection having been independently confirmed twice at two different sites (Rochester and KGI). 642 out of 660 interactions are novel; 18 were reported earlier. We did not detect 147 interactions previously reported in the literature corresponding to the 53 query mutants (Supplementary Table S3. We directly tested 57 of these previously reported suppressor interactions (corresponding to seven query mutant genes), chosen arbitrarily from among the 147 interactions that we did not discover, and were able to confirm by our methods only 15/57 interactions (Supplementary Table S4). These observations suggest that strain background and allele differences, copy number of the plasmids used for suppression, and/or the levels of galactose-induced gene expression obtainable by these various methods are sufficiently different such that a direct comparison is not possible. Therefore each case of validated suppression should be considered as a suppression event that is true under at least one set of conditions.

**General properties of the dosage-suppressor network**

The dosage suppressor network discovered here (dataset DS-A), containing 53 *ts* mutants and 517 suppressor genes, exhibits a large connected component (560 nodes, 656 edges) (Supplementary Fig. S2) that excludes 29 nodes and 20 edges. Of the 517 suppressor genes, 134 are of unknown function at the time of writing. Previous work[29] had examined the entire collection of dosage suppressors of essential gene mutants known at the time (dataset DS-B, in BioGrid), including 214 suppressor genes they discovered for 29 mutants (dataset DS-C). All three network datasets together (DS-ABC) comprise 2,286 dosage-suppressor interactions (Fig. 2; Cytoscape network output file: http://tinyurl.com/kmsb6nm).

We examined the functional similarity of the genes discovered through this study with those in previous studies in the following way. We mapped the genes of each of the three networks on to a curated and integrated protein-protein interaction (PPI) network[18,34,35,36,37] (see Methods), and calculated topological properties of the corresponding protein nodes in this PPI network (Supplementary Table S5). DS-A comprising 660 interactions overlapped significantly with the PPI network ($P=2.8 \times 10^{-15}$, see Tables 1, S5, and S6). This overlap, though significant, is small: 3.4% (23/660) of the dosage suppressor interactions in this work are edges in the PPI network compared with 19.5% (445/2,286, including 45 reciprocals) for DS-ABC (Fisher's exact $P < 0.0001$). The previous dosage-suppressor collection (DS-A and DS-B) also showed a statistically significant overlap with the curated PPI. The degree and clustering coefficient of the dosage suppressor gene nodes in the PPI network are not significantly different among the three DS datasets. By contrast, the values of another topological property, betweenness centrality (BC; see Methods for definition) of the nodes in DS-A or DS-C are significantly higher than the nodes in DS-B (Supplementary Table S5). Moreover, the mutant-suppressor gene pairs are not statistically enriched for co-expression or for genes with similar evolutionary age (Table 1), again suggesting functionally different mechanisms than stabilization by PPI, for which the interacting protein pairs tend to possess correlated gene expression patterns[38] and similar evolutionary ages[39]. A statistically significant overlap with PPI network implies that the mutant-suppressor pairs may belong to the same protein modules or complexes (below, we explicitly examine this question). However, the relatively small, though statistically significant, overlap of DS-A with the PPI network implies that DS-A is qualitatively different from DS-B or DS-C. Since a common mechanism of dosage suppression in the previously described suppressor networks was rescue of protein function through protein-protein contact[29], the present results open the possibility that a significant proportion of dosage suppressors discovered in this work might suppress through other mechanisms.

**Modular organization in dosage-suppressor network**

Biological networks have underlying modular sub-structures that reflect functional organization and evolutionary origins of gene products[38,40]. Because there is no unambiguous definition of a module within protein networks, we here examine three concepts of modularity that are intrinsically different. First, we examined protein complexes that are manually curated clusters obtained from physical protein-protein interaction data[41] (See Methods for the integrated PPI dataset), and determined the overlap of the 660 mutant-suppressor pairs with these complexes[31,41]. The products of 54 pairs of the 660 dosage-suppressor pairs were found within the same protein interaction complexes (binomial $P < 10^{-15}$) (Table 1). When the entire collection of 2,286 dosage-suppressor pairs known so far (DS-ABC) was so examined, 558 pairs co-occur in the same protein complexes (binomial $P < 10^{-15}$). Second, modules were obtained by computationally optimizing a modularity measure on the PPI network[42], and similar associations were observed within the computationally predicted PPI clusters (Table 1). Third, we identified modules dynamically, by sequentially removing genes from a curated PPI network constructed from the dosage suppressor pairs (see Methods), starting with the highest BC gene, and re-computing the BC of nodes in each resulting network, leading to a measure of modularity[43]. Gene pairs that predominantly lie within modules should remain connected in the PPI network longer than the average pair, while gene pairs that straddle modules should separate earlier. Throughout this iterative process the mutant-suppressor pairs in DS-A were more likely to be found within module boundaries (Wilcoxon rank-sum $P=2.26 \times 10^{-14}$; Supplementary Fig. S4) than were the randomly chosen protein pairs, even more so than the pairs in DS-C[29] ($P=1.55 \times 10^{-13}$) or those DS-B ($P=2.14 \times 10^{-10}$). That the mutant-suppressor pairs lie preferentially within module boundaries is consistent with the observed distribution of mutant-suppressor distances within the PPI network (number of edges along the shortest path in the PPI network connecting the mutant to the suppressor; see inset in Supplementary Fig. S4).

To determine the identity of the complexes within which the mutant-suppressor pairs co-occur, we culled from BioGRID database 4,632 direct PPIs, and used Netcarto module clustering algorithm[42] to generate 41 modules (Supplementary Table S7). We queried each mutant-suppressor pair discovered in DS-A for their co-occurrence in these 41 modules, and found eight such modules containing five clusters and three singletons (Fig. 3a-e). In one module (Fig. 3a), mutations in cell cycle control genes *cdc28* (a CDK), *cdc20* and *cdc16* (both anaphase promoting complex protein genes), and *cdc37* (encodes an HSP90 co-chaperone), are suppressed by several genes including a ribosomal protein gene *MRPL50,* and *MPD1* that encodes an endoplasmic reticulum chaperone interacting protein, underscoring the importance of molecular chaperones and ribosomal proteins in facilitating the suppression of point mutant alleles. In a second module (Fig. 3b), mutations in *cdc9* (DNA ligase), *tfb3* (transcription coupled DNA repair), and *pob3* (encodes a member of the FACT complex for nucleosome reorganization) are in the same module with several suppressors including *ADE2* (purine biosynthesis), *PSY3* (DNA repair-recombination), and *SSL2* (DNA repair helicase). While it is surprising that an adenine biosynthesis gene *ADE2* is a suppressor of *cdc9*, Ade2p has predicted protein-protein interactions with Cdc9p, Pob3p, Tfb3p, and Ssl2p, suggesting that the suppression of *cdc9* by *ADE2* likely has a biological basis through the DNA repair-recombination pathway. When this module is expanded by querying the yeast integrated genetic and physical interaction network[44] with members of this module, the network output provides further indirect evidence for linking the purine biosynthesis related genes to DNA repair-recombination (Supplementary Fig. S3). These results demonstrate the utility of using modules instructed by the dosage-suppressor network for novel gene-function discovery.

**Functional similarity within mutant-suppressor pairs**

Dosage suppressor genes and their corresponding suppressed mutant genes are functionally related. 120 out of 660 mutant-suppressor pairs in DS-A shared the same gold standard gene ontology (GO)[45] terms (binomial $P < 10^{-15}$), and 922 of 2286 known mutant-suppressor pairs (in DS-ABC) shared the same (binomial $P < 10^{-15}$) (Supplementary Table S8). The mutant-suppressor pairs have similar GO molecular functions ($P$ = 6.02 x $10^{-4}$ for the 660 pairs in DS-A, and $P < 10^{-15}$ for DS-ABC) and similar GO biological process (Wilcoxon rank-sum $P$ = 1.88 x $10^{-12}$ for DS-A, and $P < 10^{-15}$ for DS-ABC).

To test whether the dosage-suppressor pairs are enriched for other types of genetic interactions, we intersected these pairs with known genetic interaction pairs reported in the literature[46], and found significant enrichment ($P$ = 5.34 x $10^{-8}$) for 11 kinds of genetic interactions considered by Costanzo *et al*[46] (Supplementary Table S9). For 30 of the 660 gene pairs in DS-A (~4.5%), a previously reported genetic interaction was found to exist, including 18 pairs of dosage-suppressor interactions, one each of positive genetic interaction and phenotypic suppression, seven negative genetic interactions, 11 synthetic lethal/rescue or growth defect interactions, and three phenotypic enhancements. A large fraction (547/660) was not found to overlap with known physical or genetic interaction pairs, thus reinforcing the previous conclusion that dosage-suppressor data provide additional information on biological function of genes that are not provided by other types of genetic interaction data[29].

**Co-suppressors are functionally diverse**

We analyzed whether co-suppressors (suppressors of the same *ts* mutant) are functionally related to each other. First, suppressors in DS-A are not significantly enriched (Fisher's exact $P$ = 0.65) for paralogs: only 93 out of 517 genes in the suppressor network have at least one known paralog, as determined by the curated list of "ohnologs" (paralogs descended from whole genome duplication events)[47]. Among the 93 paralogs, we found 20 paralogous partners within mutant-suppressor sets (Table 2). We determined functional similarity between proteins by comparing their MIPS (Munich Information Center for Protein Sequences, http://www.helmholtz-muenchen.de/en/mips/services/index.html) functional catalog annotations[48]. Functional similarity, defined by the functional congruence (see Methods) of a *ts*-mutant gene with its suppressors, is significantly lower than that which would be expected for proteins having a direct physical interaction (Wilconxon rank-sum $P$=2.26 x $10^{-14}$) (Supplementary Fig. S5A). As a comparison, the functional congruence between a *ts*-mutant and its suppressors is also lower for DS-C network than that for DS-B (BioGrid), whereas that of DS-B is comparable to that in the curated PPI network (Supplementary Fig. S5A). These observations suggest that DS-A and DS-C reveal suppressors that are qualitatively different from those revealed by the focused methods used by earlier workers, which predominate in DS-B. Moreover, the functional congruence among co-suppressors of the same mutant in DS-A is comparable with the congruence between proteins that share a physical edge in the curated PPI network (Supplementary Fig. S5B). However, the co-suppressors of the same mutant for DS-A are considerably more diverse than those in DS-B or DS-C (Supplementary Fig. S5B), once again demonstrating a distinct collection of dosage-suppressors discovered here. We schematically illustrate this diversity in Supplementary Fig. S6.

**Plasticity and robustness of RNA polymerase II complex**

To test how often genes whose products are known to function within the same macromolecular complex can suppress mutations that affect products within the same or related complexes, we chose as a test bed a well studied protein machine—the RNA polymerase II (RNA Pol II) complex[49]. RNA Pol II core (12 subunits) is recruited to the promoter by the general transcription factors TBP, TFIIA and TFIIB (1 subunit each), and TFIIF (3 subunits), with the help of the SRB/mediator complex (25 subunits), to form the pre-initiation complex, following which TFIIE (2 subunits) and TFIIH (9 subunits) are recruited[50]. The SWI/SNF (11 subunits) and SAGA (22 subunits) complexes facilitate chromatin remodeling during transcription initiation[49]. Extensive genetic interaction studies among mediator complex proteins have been reported[51]. We scanned eight mutant genes, each of which encoded a defective (or had a complete loss of one) transcription initiation complex protein[52,53], for gene dosage suppression by 75 sub-complex genes (Supplementary Table S10; Supplementary Fig. S7). Six mutants were temperature sensitive (*ts*) lethal due to missense mutation (*med4, med11, tfb3, rad3, kin28, taf12*); two were *ts* deletion mutants (*rpb4−Δ,* and *taf14−Δ*). The deletion mutants had complete deletion of the respective structural genes, such that there was no possibility of expression of any remnant protein fragment[32,54].

31 out of 122 dosage suppressor-mutant interactions in DS-A reside within the respective complexes. For example, *MED11, NUT1, GAL11, ROX3, SRB5,* and *SRB7* each suppressed *med4* (Fig. 4a). By contrast, nearly three times as many mutant-suppressor interaction pairs (91/122 compared to 31/122; $P = 0.0001$ by Fisher's exact test) overlapped two separate sub-complexes (*e.g.,* suppression of *med4* by *TFB3* of TFIIH complex and *RPC10* of Pol II core complex, respectively). Suppressor interactions were specific: *e.g., TFA1* (TFIIE complex) suppressed *med4* but not *med11*—both of the latter encode mediator proteins. Similarly, *SNF11* and *SNF12* of the SWI/SNF complex each suppressed *tfb3* but not *kin28* (TFIIH); *RPO21* (Pol II core) suppressed both *tfb3* and *kin28*.

We discovered 36 suppressors of the eight RNA Pol II gene mutations, which are not known to be members of the RNA Pol II complex genes (Supplementary Table S2). The most striking example involves *taf14−Δ* (Fig. 4b), which yielded 27 suppressors, most of them encoding proteins outside the RNA Pol II complex, 12 of which are genes of unknown function. While the co-suppressor network of *taf14−Δ* was not enriched in PPI, a number of negative and positive epistatic edges connect several co-suppressors, including several genes of unknown function, to RNA Pol II complex genes. These results demonstrate the remarkable ability of this essential protein machine to function despite drastic genetic perturbations.

**Suppressors of a chromosome condensation defect illustrate molecular rewiring**

The genome wide suppressor dataset allowed us to explore suppression mechanisms quite different from that of the RNA Polymerase II findings discussed above: we provide examples wherein increased expression of genes allowed the bypass of an essential gene function by engaging *alternate* genetic pathways. Results described below show that the suppression of *smc2* mutation by at least two different suppressor genes appear to proceed through this general mechanism, including the engaging of proteins important for or control of meiosis to an otherwise mitotic cellular division cycle.

Smc2p is a DNA-binding subunit of the Smc2p/Smc4p condensin complex[55,56,57] required for sister chromatid (SC) alignment, separation, and inhibiting SC recombination during mitosis[55,58]. We identified four strong dosage suppressors of *smc2*: *UME1*, *MEK1*, *HTA2*, and *SNU66* (Fig. 5a). Strikingly, the first two are known to play mutually opposing functional roles[59,60,61]—*UME1* is a mitotically expressed gene required for the repression of a subset of meiotic genes, including those important for meiotic homologous recombination, and *MEK1* is a meiosis specific protein kinase that promotes meiotic homologous recombination by suppressing sister chromatid (SC) recombination. To provide more insights into the mechanisms of suppression, we analyzed the time course of mRNA expression by these four suppressed *smc2-8* mutant strains at the restrictive temperature and compared their global gene expression patterns with that of the mutant complemented by pGAL:SMC2 (See Methods).

Results (Fig. 5b-c) show that all four suppressors partially induce the expression of some meiosis-related genes. This led us to propose and test a simple model of *smc2* dosage suppression mechanism, in which meiosis specific genes rescue *smc2* defects in mitosis (Fig. 5d-e): *smc2* mutation, which causes a failure of chromosome condensation in mitosis, allows the initiation but not the resolution of SC recombination, thus blocking mitosis[55]. Ectopic expression of meiosis-specific genes in the suppressed strains either prevents precocious SC recombination or resolves the SC recombination intermediates allowing mitotic division to progress. This model is a particular instance of the general mechanism shown in Fig. 1a by genes *f* and *g*. Specifically, a non-essential meiotic gene module controlling recombination replaces the essential mitotic gene module controlling chromosome condensation. We tested an implication of this hypothesis by introducing 37 MORF clones (Supplementary Table S12) corresponding to two categories of genes into *smc2-8* and assaying their ability to suppress the *ts* growth defect: (a) other SMC complex genes, and (b) several meiotic recombination-promoting genes. 29/37 genes so tested suppressed *smc2-8* (examples shown in Fig. 5d). Among those that suppressed *smc2-8*, were *RAD51*, *DMC1*, and *MND1*—all three are recombination-promoting genes, although *RAD51* was a weak suppressor, and all suppressions were galactose independent. Several intron-containing meiosis-specific recombination genes, including *DMC1* and *MND1,* are expressed and spliced to the mature form at low levels in mitosis but are expressed highly and spliced efficiently in meiosis[62]. The function of *DMC1* is to promote recombination between homologous chromosomes and also to inhibit SC exchange in meiosis[63]. Dmc1p participates in a cascade of reactions activated through phosphorylation by Mek1p[60], which we have found also to be a suppressor of *smc2*. These results support the idea that at least one mechanism of suppression of *smc2* is through mitotic expression of meiotic recombination genes, which are expected to prevent the formation of or to promote the resolution of sister-chromatid junctions that occur at high frequency in *smc2* mitosis.

**A link between chromosome condensation-cohesion pathways and Huntington's disease**

A query of combined PPI and genetic interaction databases with six genes—*smc2* and its five suppressors *UME1, BNA5, SCC4, SMC1,* and *SMC3*—generated an interaction network of 26 genes (including *SCC2*) containing three functional modules (enrichment significance, all Benjamini-Hochberg (B-H) corrected: chromatin remodeling complex, $P = 2.37 \times 10^{-5}$; sister chromatid cohesion complex, $P = 1.38 \times 10^{-7}$; and chromosome condensation, $P = 6.4 \times 10^{-13}$) (Fig. 6a). The members of each module exhibit extensive PPI and genetic interactions with members of the other two modules. We noticed that *SCC2* encodes a member of the HEAT repeat proteins, which include Huntingtin (Htt)—a protein with expanded poly-glutamine residues in the N-terminal region that polymerizes to an insoluble aggregate (plaque) in certain brain cells of Huntington's disease patients[64]. Moreover, *BNA5* is a suppressor that linked two functional modules (Fig. 6a). A previous study[65] had identified 28 deletion mutations that suppressed Htt toxicity in yeast, which included *ume1Δ* and *bna4Δ,* both of which have mammalian homologs. The transcriptional repressor Ume1p requires for its activity Rpd3p[59], a part of the histone deacetylase (HDAC) complex. The inhibition of Rpd3 HDAC complex by Ume1p increases the toxicity of mutant human Htt in yeast, where Htt toxicity was related also to the function of *BNA4* and *BNA5*[66]. The latter two genes encode two successive enzymes in the NAD biosynthetic pathway from kynurenine, and the *BNA4* product is a target for therapy against Huntington's disease[67,68,69]. These results are consistent with the hypothesis that Htt toxicity is related to the chromosome condensation-cohesion processes, and might be influenced by homologous recombination pathways. We tested one prediction of this hypothesis by attempting to suppress Htt toxicity by *SMC2*, suppressors of *smc2-8*, and cohesin/condensin genes (Fig. 6b): 9 of 14 genes so tested suppressed Htt toxicity (*SMC1, SMC3, SMC4, SCC4, MEK1, UME1, DMC1, HTA2,* and *MND1* suppressed; *SMC2, SCC1, SCC3, CWC24* and *RAD51* did not).

**Discussion**

A genome wide network of dosage suppressors allowed us to explore functional relatedness among the co-suppressors. Mapping of the mutant-suppressor pairs on a PPI network revealed boundaries of topologically defined protein modules. By examining specific protein modules through suppressors of selected RNA Pol II gene mutants, we found that high expression of specific component proteins within large protein assemblies can functionally replace the absence or the reduction of specific essential components. These latter results underscore the importance of including systematic dosage suppression data in deriving systems-level models of large protein complexes and their pathways of self-assembly.

In principle, dosage-suppressor interactions may involve high affinity and high probability protein-protein interactions that enable the system to return to the original state by positive and/or negative feedback effects (buffering interactions). In one scenario, over-produced suppressor products stabilize the corresponding thermo-sensitive missense mutant protein by direct interaction, through recruiting functionally competent folding intermediates from a distribution of folded states. By contrast, suppressors of deletion mutations, such as those of *rpb4−Δ* and *taf14−Δ*, in which the entire coding frames were deleted, cannot possibly exert their effects by stabilization through direct protein-protein interaction. Surprisingly, deletion *ts* alleles had a similar average frequency of suppressors (121 interactions with 10 mutants) as did missense *ts* alleles (539 interactions with 43 mutants). This suggests that heat sensitive biological processes (for *ts* deletion mutations) are as suppressible as processes catalyzed by individual heat-sensitive proteins (for *ts* missense mutations). Thus, the suppression mechanisms of

*rpb4−Δ* by *RPB3* (RNA Pol II core), and by *ROX3* (mediator) likely involve direct or indirect functional replacement of Rpb4p. One mechanism might be the stabilization of the RNA Pol II preinitiation complex by the over-expression of one of the component proteins, such that the preinitiation complex rendered heat labile by *rpb4−Δ* is made more robust by an overabundance of Rpb3p or Rox3p. A similar argument holds for the suppression of *taf14−Δ* by *TAF2, TAF13,* and *TAF10* (all TFIID), *TFB3* (TFIIH), *SRB7, PGD1, CSE2* (all mediator/SRB), *SPT3* (SAGA)*,* and by *SNF5* and SNF6 (both SWI/SNF). Such alternate functional replacements within and between protein complexes reflect a high degree of compositional plasticity, and might also imply alternate pathways of assembly of multi-protein complexes. These observations are generally consistent with the recent observation that RNA Pol II open complexes can be reactivated by the TFIIE complex through stabilizing effects on relatively unstructured domains on mediator proteins[70]. Such "Intrinsically Disordered Regions" serve to functionally assemble RNA pol II complex subunit proteins[54,71] and thus provide a high degree of modular functionality.

Although 45 of 53 suppressed mutant alleles are missense mutations, and several suppressors encode protein folding or processing enzymes such as chaperones or heat shock proteins (*HSC82*, *HSP32*, and *CCT6*) or chaperone interacting protein (*MPD1)*, there is no significant statistical enrichment for these classes of proteins in the dataset produced in this work. By contrast, there is a significant enrichment for ribosomal proteins in our dataset (B-H corrected $P = 2.64 \times 10^{-4}$) as in the full set of known dosage suppressors (B-H corrected $P = 4.5 \times 10^{-6}$). The suppression of *cdc37* (a co-chaperone mutant) by *RPS18A*, *RPL25*, and *RPS24B* is consistent with the possibility that some ribosomal proteins may have weak chaperone-like activity[72].

Some suppressors of *smc2-8* appear to act through direct protein-protein interaction with the mutant protein. For example, *SMC1* and *SMC3*, required for SC cohesion, but not the condensin gene *SMC4*, can suppress *smc2-8* (Fig. 5d). Protein-protein interaction between the mutant Smc2-8p and Smc1p/Smc3p cohesin complex might be able to stabilize misfolded Smc2-8p, whereas direct interaction between Smc2-8p and Smc4p, both members of the condensing complex, cannot do so. A recent report indicates that Smc2p homolog from *S. pombe* interacts with the *S. pombe* histone H2A and H2A.Z proteins in recruiting the condensin proteins to mitotic chromosomes[73]. Since *S. cerevisiae HTA2* encodes a homolog of *S. pombe H2A* gene family, it is possible that Hta2p also recruits Smc2p to the nucleosome in an analogous manner. If true, at least a part of the suppression mechanism by *HTA2* might be explained by the stabilization of mutant Smc2p through direct protein-protein interaction with Hta2p. Note that *snu66* and *hta2* mutations genetically interact[46], suggesting that the mechanisms of suppression of *smc2-8* by the two corresponding wild type genes might indeed be mutually related. Snu66p interacts with Snu40p[46,74], and Snu40p in its turn interacts with proteins involved in chromosome condensation and cohesion, including Scc3p [75]. Scc3p is part of the Smc1p-Smc3p-Scc1p-Scc3p cohesin complex, which interacts with the Smc2p-Smc4p condensin complex, and the latter is stabilized by Scc2p-Scc4p complex. Therefore, one mechanism might be the stabilization and recruitment of thermosensitive Smc2-8p subunits into the condensin complex by direct protein-protein interaction with a cohesin complex protein that includes Scc4p. The suppression of *smc2-8* by *SCC4* is consistent with this mechanism.

Dosage suppression by rewiring, in contrast to that by direct PPI, may involve low affinity and/or low probability interactions that illustrate alternative—redundant or degenerate—pathways. These pathways of suppression appear to decouple physical interaction modules from the modules of functional activities, and the flexible interaction edges rearrange the functional modules to buffer the effects of genetic and environmental perturbations. We presented evidence that a mitotic chromosome condensation defect can be bypassed by ectopic expression of a series of meiotic genes involved in controlling sister chromatid recombination and resolution of recombination intermediates, suggesting that a defective biological process can be bypassed by alternate biological processes that are not engaged in cells during normal function. We proposed and tested a model in which the mitotic blockage by the *smc2* mutation could be bypassed by *UME1*, *MEK1*, *HTA2*, and *SNU66*, by either preventing SC recombination initiation or augmenting SC recombination resolution; the predicted model was tested by further findings that several meiotic recombination genes *DMC1*, *MND1*, and *RAD51* also suppressed *smc2*. Since *SNU66*, a strong suppressor of *smc2*, encodes a splicing component, it is possible that the over-expression of Snu66p augments the splicing of DMC1 and MND1 pre-mRNAs in mitotic cells.

Suppressor interactions that connect *BNA4/5* to *UME1*, *SCC2*, and *SMC2*, point to a possible mode of action of the mutant Huntingtin protein involving the chromosome condensation/cohesion process. As a test of this hypothesis, we examined the ability of a set of suppressors of *smc2-8* mutation (including *SMC2*) to also suppress 103Q toxicity, and discovered that several cohesin/condensin genes, and a few homologous recombination genes including *MND1*, are able to do so. Interference by Htt of the normal sister-chromatid cohesion process could potentially increase somatic recombination at CAG repeats, and thus might underlie the observed variable penetrance of high poly-glutamine expansion alleles on disease onset age in humans[76,77]. Consistent with this speculation, a recent report has linked a network of ploidy control genes in yeast to Htt toxicity[78].

An unresolved question remains as to the extent of genetic robustness that is hardwired in the genome, which was first brought to the fore by the investigation of mutational effects on the lysis-lysogny decision circuit of bacteriophage lambda[79]. In this work, mechanisms of suppression of a defect in chromosome condensation revealed insights on the potential of unrelated genes that could be brought to bear on solving problems associated with defective cellular processes. It is conceivable that yeasts in nature, and organisms in general, depend on the rewiring of gene regulatory circuits to find new solutions to essential cellular processes during evolution under selective pressure, as observed in this work that meiotic genes relieve mitosis blockage. Such a possibility was investigated earlier in evolved yeast strains with a deletion in an essential gene (*myo1*), where aneuploidy and large-scale variation in the transcriptome were associated with survival[27]. While aneuploidy was a recurrent theme observed in that work, it was also estimated that the number of available genetic solutions to a lethal perturbation might be limited. Gene redundancy[80], and promiscuity of gene function[30,81] both contribute to genetic robustness. While the core set of essential genes might impede evolvability[82], modular rewiring may in principle overcome this barrier[83]. Our finding that nearly six times as many genes can suppress 53 deleterious mutations indicates a high degree of robustness built into the genome, and illustrates potential pathways for rewiring of the genome. It is conceivable that a deleterious mutation in an essential gene, leading to growth arrest, is followed by genomic changes that are often observed in stationary phase cells[27,84,85,86,87]; these changes could in principle activate suppressor pathways to restore viability and provide adaptation.

The network of dosage suppressors of essential gene mutants is analogous to the network of genes that could potentially bypass, if aberrantly expressed, a drug target gene function (*e.g.*, of a cancer-essential gene) for tumor cell survival. Such a network for a cancer cell is the equivalent of potential pathways for developing resistance to cancer chemotherapy, or, analogously, for evolving independence from the checkpoints that ensure non-proliferative growth, which evidently occurs frequently during the development of cancers.

**Materials and methods**

**MORF plasmids**   The movable open reading frame (MORF) library[33] containing 5,871 2μ plasmids with galactose inducible promoter and a *URA3* selectable marker were divided into 16 pools. Each pool, representing approximately 384 plasmids, was grown in 96 deep-well plates, pooled, and plasmid DNA samples were isolated for transformation.

**Yeast strains, media and transformation**   Temperature sensitive lethal *Saccharomyces cerevisiae* strains had specific mutations in BY4741 background (*MATa his3-Δ1 leu2-Δ0 met15-Δ0 ura3-Δ0*); point mutants were provided by Dr. Charlie Boone (University of Toronto) and *ts* deletion mutants were screened and selected from the deletion mutant library (OpenBiosystem). For each mutant, the range of growth and the threshold of non-permissive temperature on both inducing (+ galactose) and non-inducing (- galactose) conditions were determined. Yeast strains were grown in yeast complete media containing 1% raffinose, transformed with 1μg of each MORF plasmid pool, and plated at permissive temperature on synthetic defined medium lacking uracil with 1% raffinose. The transformants from 16 plates were pooled and selected at the restrictive temperature for that particular *ts* allele on complete and synthetic media containing either 2% glucose (repression) or 2% galactose (induction) (see Supplementary Table S1) for a list of restrictive temperatures corresponding to the *ts* alleles).   The threshold restrictive temperature that cuts off the growth of each individual allele with or without (vector control) the candidate suppressor plasmid was determined for each suppressor by incubating identical multiple-replicate plates at a range of temperatures spanning at least $\pm 2^0 C$ around the restrictive temperature for that allele. Transformants in each mutant strain were selected for growth above the respective restrictive temperature characteristic for the corresponding mutant strain containing the empty vector plasmid. Strains for testing suppression of Htt toxicity were in W303 background[88] (*can1-100, his3-11,15::FLAGhtt103Q-CFP,leu2-3,112trp1-1,ura3-1, ade2-1*; and *can1-100, his3-11,15::FLAGhtt25Q-CFP, leu2-3,112trp1-1,ura3-1,ade2-1*).

**Suppressor identification and confirmation** Suppressor candidate hits were identified by microarray hybridization of isolated plasmid DNA from colonies growing above the restrictive temperature for the respective *ts* allele as follows: Approximately 300 colonies were picked from selection plates at restrictive temperature, grown in 96 deep-well plates. The cells were pooled and plasmid DNA isolated using Cycle Pure Kit from Omega Bio-Tek, and labeled with Cy3 dye, whereas the pooled DNA of the MORF library was labeled with Cy5 by PCR amplification using two flanking primers (5'GGACCTTGAAAAAGAACTTC3', 5'CCTCTATACTTTAACGTCAAGG3'). Labeled probes were hybridized to spotted microarrays (UHN Microarray; containing > 95% of all ORFs) at 65 $^{\circ}$C for 16 h. Microarrays were scanned in Bio-Rad VersArray Chip Reader and the data were analyzed using ScanArray express software 3.0 (Perkin Elmer). For a limited number of suppressors, we purified the DNA from colonies growing on selection plates with raffinose and galactose, transformed into *E. coli*, re-isolated the corresponding plasmid DNA samples, PCR amplified the ORF off the MORF plasmids, and sequenced the DNA. In all cases, the suppressor MORFs identified by sequencing were identical to the MORFs identified by microarray hybridization. The putative suppressor genes identified by either microarray or by direct sequencing were retransformed individually to the respective mutant strains, their ability to suppress the mutants confirmed individually on at least three separate transformed colonies by isolating single colonies and testing on inducing or non-inducing plates at a range of temperatures above the growth cutoff temperature of the corresponding unsuppressed mutant, dependence of suppression on the introduced MORF plasmid was confirmed for each transformant on plates containing 5-fluroorotic acid, and for those transformants passing all above tests the titration-spotting test was carried out for final confirmation/quantification of suppressor strength. The background strain was always compared on the same plate with the candidate-suppressed strains at a range of temperatures spanning at least $1^0$C over the corresponding threshold temperature for growth of the given mutant. The extent of suppression was subsequently quantified through spotting of serial dilutions of each mutant/suppressor culture under both inducing and noninducing conditions. A subset of the final list of suppressors was again confirmed by sequencing. The negative control for each suppressor was the corresponding mutant strain carrying the empty MORF vector (BG1776). Positive control plasmids were the complementing genes under pGAL control, except for six mutants (*cdc13, cdc4, cdc15, cdc35, cdc48,* and *abd1*) that did not have the appropriate positive controls because either the over-expression of the corresponding MORF plasmids was lethal (*CDC13*, and *CDC48*) or they were absent in the MORF library (*CDC4*, *CDC15*, *CDC35*, and *ABD1*). The strength of suppression for each suppressor was normalized to the growth of the diluted spots against that of the corresponding vector control (BG1776) strain on the same plate on adjacent rows (Supplementary Table S2).

**Protein detection**  Cultures were grown in repressing and inducing media, and whole cell extracts were prepared by the bead beating method in yeast lysis buffer (25mM Hepes-NaOH pH 7.5, 10mM NaCl, 1mM EDTA, 0.1% Triton X-100) containing

EDTA–free complete protease inhibitor tablet (Roche). Proteins were detected by Western Blotting, probed by anti-HA antibodies (Covance) using standard methods.

**Liquid growth assay** Growth curves in liquid media along with maximal growth rates were determined using a Bioscreen C Automated Growth Curves Analysis System (Growth Curves USA). The suppressed strains were grown in 200 μl of S-URA with 1% raffinose and 2% galactose at various temperatures in 96-well plates. The optical density (OD) was measured at 600 nm every 30 minutes for 48 hours of growth.

**RNA methods** Strains were grown in 5 mL S-URA media containing 2% raffinose for 24 hours at 28°C. Samples were diluted to OD 0.02 in 150 mL S-URA media containing 2% raffinose and 2% galactose, and then grown overnight at 25°C with agitation. At OD 0.1, cultures were shifted to 34.5°C with continuous agitation, and samples were harvested at 0, 45, 90 and 180 min by centrifugation. Total RNA was isolated using a hot phenol method [89], followed by 2 chloroform extractions, RNA precipitated by addition of 1/10$^{th}$ volume of NaOAc pH 5.2, and 2.5 volumes of 100% ethanol and the pellet dissolved in 50μL ddH$_2$O. Total RNA was deproteinized again after DNase treatment and resuspended in 10 μL of ddH$_2$O, and quantified by a nanodrop spectrophotometer.

**Gene expression measurements** 1 μg total RNA samples from two individual biological replicates, four time points each (0, 45, 90 and 180 min), of *smc2-8* mutant strains harboring pGAL:SMC2, pGAL:UME1, pGAL:MEK1, pGAL:HTA2, pGAL:SNU66, and the negative control MORF plasmid (pGAL:negative, BG1766), were reverse transcribed, hybridized to Affymetrix Yeast Genome S98 arrays and scanned with Affymetrix GeneChip Scanner 3000. The microarray data were analyzed with GeneSpring 6.2 software, and were deposited in Gene Expression Omnibus (accession number GSE24266). Microarray expression levels were verified for 7 reference genes by quantitative RT-PCR using samples of a third biological replicate (Supplementary Fig. S8).

**Microarray Data Analysis** The Affymetrix array data were processed using the Robust Multi-array Analysis (RMA) as described previously[90]. A log scale, linear additive model represented the perfect match and mismatch data. For each experiment (time point or condition), the RMA analysis produced one numerical estimate of expression for every probe on the chip (two replicates for each treatment and time point). We combined the replicates using a median based normalization (given microarray replicates 1 and 2, determined the median intensities $m_1$ and $m_2$ of microarray 1 and microarray 2 respectively; adjusted the values of microarray 2 by adding $m_1 - m_2$ to the intensities of microarray 2) to produce an average of the adjusted intensities. For each time point and treatment we produced one intensity measurement for each probe. These numbers were used to find ratios of fold change from one time point to the next. For each chip, a background noise intensity measure was formed using the average intensity of the SPACER probes which act as a set of negative controls for the chip – if for a particular probe, the intensity level at time points $t_1$ and $t_2$ was below the background noise level at $t_1$ and $t_2$, then we assumed the probe was expressing at background noise level and the fold change was set to 1. We clustered genes that were significantly differentially expressed in at least one time point.

**Quantitative RT-PCR** DNase treated RNA was reverse transcribed in 25 μL RT reaction mixtures (1x First Strand Buffer, 0.02 μg random hexamers, 0.01 M DTT, 0.5 mM dNTP mix, 0.6 U RNaseOUT, containing 3 U Superscript II Reverse Transcriptase) for 2 h at 37 ºC, followed by heat inactivation at 100ºC for 5 min and quick chilling on ice. A standard curve was generated for each gene target starting with 0.5 µg RNA, and four successive 2-fold serial dilutions. The cDNA templates generated by reverse transcription was used for quantitative RT-PCR in Applied Biosystems 7500 FAST Real Time PCR system. The PCR mix constitutes 20 µL buffer containing 1X FAST SYBR Green Master Mix and 0.2 µM forward and reverse primers (IDT), with the PCR conditions: [95°C → 20 sec] HOLD, [95°C 3 sec → 60°C 30 sec] 40 times). The following primers were used for real time PCR of seven genes:

Table: List of Primers for Real Time Quantitative PCR

| Primer | Sequence |
|---|---|
| OPT2 Forward Primer | GGG CTT TGA ATT TGT GGG CCA TGA |
| OPT2 Reverse Primer | TCA TAA TCG TCG AGC GCC CTG TAA |
| SMC2 Forward Primer | AAC TTG TGC CGG AGG TAG GCT ATT |
| SMC2 Reverse Primer | GCC AAT TCA ACT TTC CCA GGA GCA |
| PHO5 Forward Primer | AGA CAT GCT CGT GAC TTC TTG GCT |
| PHO5 Reverse Primer | AAG CAC TCA AAG TGT TGG CAC CAG |
| CYC7 Forward Primer | AGT ACG GGA TTC AAA CCA GGC TCT |
| CYC7 Reverse Primer | GTC CAA CTT TGT TAG GAC CAC CCT |
| GRE1 Forward Primer | ACT GGT GGT GGC ACT TAT ACC CAA |
| GRE1 Reverse Primer | TGG TAG CGG TTA CTT TGA GCA CCT |
| SIP18 Forward Primer | AGG GAA AGA ACG CCA AAT CCT CCA |
| SIP18 Reverse Primer | CAA TCG TTC GCA ATT CCT CTG CCA |
| FIT1 Forward Primer | TGC CCA ATC TGT TCG TAC CCA TGA |
| FIT1 Reverse Primer | ACC AGC GGT AGT GGT TTG AAC TCT |

**Datasets**

The full Dosage Suppressor (DS) dataset (DS-ABC, see text), consisting of data reported here combined with other available data, was culled from the following sources: Magtanong et al.[29], the latest BioGRID version 3.1.78 [31], and this work (660 dosage suppressor interactions between 53 mutant ORFs and 517 suppressor ORFs). The full DS dataset contains 2286 interactions between 454 mutant ORFS and 1284 suppressor ORFS, and includes 60 reciprocal interactions, resulting in 2226 unique interactions. The PPI network was constructed from BioGRID version 3.1.78 [31] that contains 6,614 ORFs of which 5,955 nodes are connected by physical interaction edges. The 99,866 physical interactions produced 60,143 unique edges among 5,728 nodes. This network was curated to filter out indirect physical interactions and retain only the direct physical interaction data from eight types of experiments including Biochemical activity, co-crystal structures and reconstituted complexes, PCA, protein-peptide interactions, two-hybrid, far western and FRET. The resulting direct PPI network, containing a total of 20,034 unique interactions between 4683 nodes, was used for studying the enrichment of PPI interactions in the dosage-suppressor (DS) network (Supplementary Table S6). Genetic interaction (GI) data was downloaded using the latest BioGRID version 3.1.78[31], which contains eleven kinds of genetic interactions, namely, Dosage growth defect, Lethality, Dosage Rescue, Negative Genetic, Phenotypic Enhancement, Phenotypic Suppression, Positive Genetic, Synthetic Growth Defect, Synthetic Haplo-insufficiency, Synthetic Lethality, and Synthetic Rescue. In all, the GI dataset includes 16898 interactions between 5411 nodes. The Stanford Microarray Database was used for inferring co-expression between yeast ORFs at a correlation coefficient cut-off of ±0.5. The resultant co-expression network includes 623,224 unique edges between 5155 nodes from total 643 experiments reported by two groups[91,92]. Genome-wide protein complex data was inferred by combining the Curated Yeast Complexes (CYC2008), a comprehensive catalogue of manually curated 408 heteromeric protein complexes in *S. cerevisiae* with 400 complexes in the Annotated Yeast High Throughput (YHTP) complexes derived from high-throughput Tandem Affinity Purification/Mass Spectrometry (TAP/MS) studies[41], and 72,016 pairs of indirect physical interactions from Affinity Capture, Co-Fractionation, Co-purification and Co-localization experiments (BioGRID version 3.1.78). The Pfam domains for many yeast proteins have been assigned one of four age groups ABE, AE/BE, E and F depending on their taxonomic distribution among archaea (A), bacteria (B), eukaryote (E) and fungi (F)[93]. This dataset was used to analyze whether DS pairs were likely to belong to the same age group. Both the DS datasets were tested for significant overlap with computationally predicted modules. For this, 41 Yeast Louvain modules were identified in the Yeast direct PPI network by using the NetCarto algorithm[42]. In addition, Markovian clusters were identified using the MCL-MLR clustering method [94] at an inflation value of 2.4.

**Paralog identification** The list of paralogs (554 gene pairs) was described before[47], and includes 457 pairs previously found[95]. Of the 1,108 paralogous genes, 1,001 were represented in the MORF library. To test for significant enrichment of paralogs, we performed a Fisher's exact test using the 5,829 testable ORFs as the baseline.

**Network properties** Betweenness centrality (BC) (fraction of shortest paths through a node) was $B'_i = \sum_{\text{all pairs}} b_i$, where $b_i$ is the ratio of the number of shortest paths between a pair of nodes in the network that pass through node *i*. BC was scaled as $B_i = \dfrac{2B'_i}{(n-1)(n-2)}$, where $n$ is the number of nodes in the network[96]. Clustering coefficient (ratio of the actual number of degrees of a node to the possible degrees given a node's neighbors), and shortest path between pairs were computed using the MATLAB Boost Graph Library toolset.

**Functional congruence assessment**    Functional gene annotations were derived from the MIPS FunCat database[48]. 449 of the 642 unique suppressors were annotated genes. Un-annotated genes (class '99') were excluded. Here, functional congruence between two genes is defined as the extent of overlap between their respective MIPS annotations. Because MIPS functional annotations are hierarchical categories, we studied the congruence of MIPS annotations based on the 1$^{st}$ category alone, 1$^{st}$ and 2$^{nd}$, 1$^{st}$, 2$^{nd}$, and 3$^{rd}$, etc. categories, down to congruence in all 5 categories. Because genes can have several MIPS annotations, we compiled the possible pairs of categories for each pair of genes and calculated the fraction of pairs of annotations that agreed, divided by the number of possible pairs of annotations. Because many annotations do not include details up to the fifth functional categorization, we devised rules to match annotation strings of different length. For example, when matching the functional MIPS annotation "01.01.05.01.02" (degradation of polyamines) with "01.01.06" (metabolism of the aspartate family), we pad annotations such that the previous example would result in a match at level 1, a match at level 2, but not at level 3 and beyond. In other words, for incomplete annotation the omitted part is assumed to be different from that of any other annotations. The functional congruence of two genes at level $n$ is the fraction of annotations of these genes that are identical up to level $n$. By design, the functional congruence of a pair at level $n$ is larger or equal to the functional congruence of that pair at level *n+1*.

## Acknowledgements

We thank Dr. C. Boone (U. Toronto) for discussion, for sharing strains and their suppressor data before publication, Dr. Susan Lindquist (Whitehead Institute, Cambridge, MA) for providing strains with the human Htt genes, B. Kraynack and K. Svay for initial work in suppressor discovery, and C. Cordova, R. Vasudevan, V. Padmakumar, K. Bheda, K. Datta, A. Kamra, D. Lev, and M. Pollock for technical assistance, and Dr. C. Murre (UCSD) for comments on an earlier version of the manuscript. This work was supported by grants from the National Science Foundation (Frontiers in Integrative Biological Research #0527023 to D. J. G, E. J. G., E. M. P., C. A., A. Rav. and A. Ray; #0523643 to A. Rav. and A. Ray; #0941078 to A. Ray), the National Institutes of Health (# 1R01GM084881-01 to A. Ray) and CHDI Foundation (to A. Ray). The authors have no conflict of interest.

## Contributions

B.P., Y.K., A.W.S., G.Y. J.F., R.V., A.H., B.Ø., A.B., B.M., and J.T. conducted experiments and analyzed data, A.K. conducted experiments; B.P., Y.K., A.W.S., J.F., R.V., A.H., B.Ø., A.B., B.M., N.S.B., E.J.G., D.J.G., G.Y., A.Rav., C.A., E.M.P, and A.Ray designed experiments and analyzed data; B.P., A.W.S., G.Y., J.F., A.H., B.Ø., A.B., D.J.G., A.Rav., C.A., and A.Ray wrote the manuscript; E.J.G., D.J.G., A.Rav., C.A., E.M.P., and A.Ray procured funding.

**Figure legends**

**Fig. 1:** Dosage suppression of essential gene mutations. (a) X and Y are cellular processes, of which X but not Y is essential for viability. X normally requires the function of genes *a-e*. Gene *e* is regulated (blue arrows) by a complex of proteins (black lines, PPI) encoded by genes *a-d*. Y is regulated by *g* and *f*. If *a* is mutated to a conditional allele (red), over-expression of either *b* or *c* (but not *d*) may stabilize sufficient properly folded mutant **a** proteins, causing suppression (green arrows). If gene *a* deletion causes *ts* lethality, and if over-expression of *b*, *c,* or *d* causes suppression, then the structural modularity of components within the complex enables suppression. If the over-expression of *f* and/or *g* also suppresses *a* and/or *e*, then genes *f* and/or *g* can also control process X through a promiscuous mechanism (dashed arrow). Suppression by *f* and *g* reveals functional modularity. (b) The strategy for isolating dosage suppressors of *ts* lethal mutations. See Methods for details. (c) A few examples of suppressors of different strengths. A *ts* allele of *cdc36* is viable at $25^0$C, but fails to grow above $33^0$C, is complemented by pGAL:CDC36 at $35^0$C or below under all tested conditions (galactose independent), and is suppressed by *MATα2* (strong, grade 4), *MTH1* (medium-weak, grade 2), *CCL1* (strong, grade 5), and *ASF2* (strong, grade 5). The latter two genes exhibit galactose independent suppression. YNL324W is a very weak suppressor (grade 1, reproducible). pGAL-negative is the vector control. (d) Summary of the screen. Most suppressors were effective on galactose but not on glucose (Tables S2); a few exceptional suppressors are galactose independent (see above), presumably because these expressed detectable quantities of MORF-encoded protein even on glucose (see Supplementary Fig. S1 for examples).

**Fig. 2:** Dosage suppressor interaction network in yeast. This network contains 454 *ts* lethal mutations that are suppressed by 1,284 genes, representing 2,286 interactions. Nodes are either essential gene mutations or their suppressors, and the directed edges (color and thickness weighted by edge-betweenness centrality) are dosage suppressor interactions. Many suppressors are likely specific to the experimental conditions adopted here, and at least a subset may be allele specific. Alternate experimental methods may reveal additional suppressors. The data contain 660 novel interactions reported here, added to those listed in BioGrid and those reported by Magtanong *et al.*[29].

**Fig. 3:** Mutant-suppressor pairs are enriched within protein modules. Five computationally predicted modules in which specific mutant-suppressor pairs are statistically enriched are shown. The Netcarto algorithm returned a total of 41 modules for the 660 ORF pairs in dataset A. Of these, 71 pairs of ORFs lie within modules (Supplementary Table S7). Node color: purple, mutant genes; all remaining nodes are suppressors, arbitrarily colored according to their functional annotations. Edge colors: green, genetic interaction; gray, dosage suppressor interaction; purple, co-expression; yellow, shared protein domains; brown, physical interaction.

**Fig. 4:** Dosage suppressors of RNA Pol II mutants. (a) Suppression of RNA Pol II mutants. We chose eight *ts* mutant genes (RNA Pol II core complex: *rpb4-D*; mediator complex: *med4* and *med11*; TFIID: *taf12* and *taf14-D*; TFIIH: *rad3*, *tfb3* and *kin28*) and scanned these for dosage suppression by 75 RNA polymerase II complex genes. Dosage

suppressors (at the base of each arrow) of mutant proteins (at the head of each arrow) are indicated. Not all sub-complex proteins are shown. Arrow denotes dosage suppression. Example titration results are in Supplementary Fig. S7. (b) Co-suppression network of *taf14-Δ* as derived from genome wide dosage suppressor screen. Edge color: yellow, dosage suppression; green, genetic interaction; blue, PPI.

**Fig. 5:** Network rewiring as a mechanism of suppression. (a) Cell cycle checkpoint arrest by *smc2-8* at the restrictive temperature is suppressed by galactose-induced expression of *UME1*, *MEK1*, *HTA2*, and *SNU66*. (b) Growth curves of the suppressed strains. *Smc2-8* strains harboring pGAL:SMC2, pGAL:UME1, pGAL:MEK1, pGAL:HTA2, pGAL:SNU66, and the negative control MORF plasmid (pGAL:negative, BG1766), respectively. Samples taken from these cultures were used in gene expression profiling. (c) Heat map of differentially expressed genes. Normalized $\log_2$ transformed mRNA levels of genes were re-normalized against corresponding expression levels in the negative control strain, the resulting expression ratios were filtered for signals above $\pm 2\sigma$ and hierarchically clustered. The first four time points are $\log_2$ ratios of expression values of *smc2-8*/pGAL:SMC2 to that of *smc2-8*/pGAL negative control plasmid. The remaining lanes are $\log_2$ ratios of expression values of *smc2-8*/pGAL:UME1, *smc2-8*/pGAL:MEK1, *smc2-8*/pGAL:HTA2, and *smc2-8*/pGAL:SNU66, respectively, to that of *smc2-8*/pGAL:SMC2. (d) Results of a focused screen for additional *smc2-8* suppressors. Only a few suppressors of various strengths are shown. (e) Mechanisms of *smc2-8* suppression deduced from the phenotype and suppression network.

**Fig. 6:** Dosage suppressor network illuminates Huntington's disease mechanisms. (a) Modules discovered by querying combined genetic and physical interaction network of yeast with suppressors of *smc2*, showing a link through *BNA5*. (b) Example results of a test that suppressors of chromosome cohesion/condensation defects in *smc2-8* mutant also suppress expanded poly-Q toxicity of the Huntington's disease protein Htt. Yeast strains carrying a chromosomal copy of either normal (25Q) or expanded (103Q) allele of the human Htt N-terminal fragment, expressed through the pGAL promoter, were transformed with MORF plasmids and tested for suppression of 103Q toxicity. Strains that express toxicity fail to grow on media containing galactose. Suppression of toxic Htt by *HTA1* was reported before[97], so served as a positive control. Here, *MND1*, *SMC1*, and *SMC3* suppressed Htt toxicity but *SMC2* did not. See text for complete results.

.

**Table 1**
**Similarities between mutant genes and their suppressors**

| Number of mutant-suppressor gene pairs | Among dosage suppressor pairs (total 660) discovered in this work | | Among all known dosage suppressor pairs (total, 2286) | |
|---|---|---|---|---|
| | Number | Enrichment significance (Binomial one-sided test $P$) | Number | Enrichment significance (Binomial one-sided test $P$) |
| With direct PPI | 23 | $2.8 \times 10^{-15}$ | 445 | $<10^{-15}$ |
| Co-located in the same protein complex | 54 | $<10^{-15}$ | 558 | $<10^{-15}$ |
| Co-located in the same computationally predicted protein module[#] | 71 | $8.6 \times 10^{-6}$ | 557 | $<10^{-15}$ |
| Co-located within the same PPI cluster[##] | 12 | $2.4 \times 10^{-5}$ | 232 | $<10^{-15}$ |
| In which both genes are co-expressed | 10 | 0.263097* | 144 | $3.1 \times 10^{-12}$ |
| In which both genes are of similar evolutionary age | 159 | 0.06422174* | 641 | $1.8 \times 10^{-10}$ |

*No statistically significant enrichment
[#]Co-located in one of 41 Louvain modules, computed by the Netcarto algorithm [42]
[##]Co-located in Markovian clusters, computed by MCL-MLR clustering [94].

**Table 2**

**Paralogous pairs of dosage suppressors**

| Paralog 1 | Mutation(s) suppressed by paralog 1 | Paralog 2 | Mutation(s) suppressed by paralog 2 |
|---|---|---|---|
| YBR181C | *cdc24, cdc15* | YPL090C | *cdc24, cdc15, poc4* |
| YCR073W-A | *cdc26* | YNR034W | *cdc26* |
| YDR099W | *cdc25* | YER177W | *cdc25* |
| YDR471W | *pol5* | YHR010W | *pol5, cdc25* |
| YER056C-A | *cdc48* | YIL052C | *cdc48* |
| YER131W | *vrp1* | YGL189C | *vrp1* |
| YFR023W | *smc2* | YHR015W | *smc2* |
| YJL177W | *cdc48* | YKL180W | *cdc48, cdc13* |
| YLL062C | *taf14* | YPL273W | *taf14, cdc13* |
| YLR029C | *cdc24, cdc26* | YMR121C | *cdc24* |

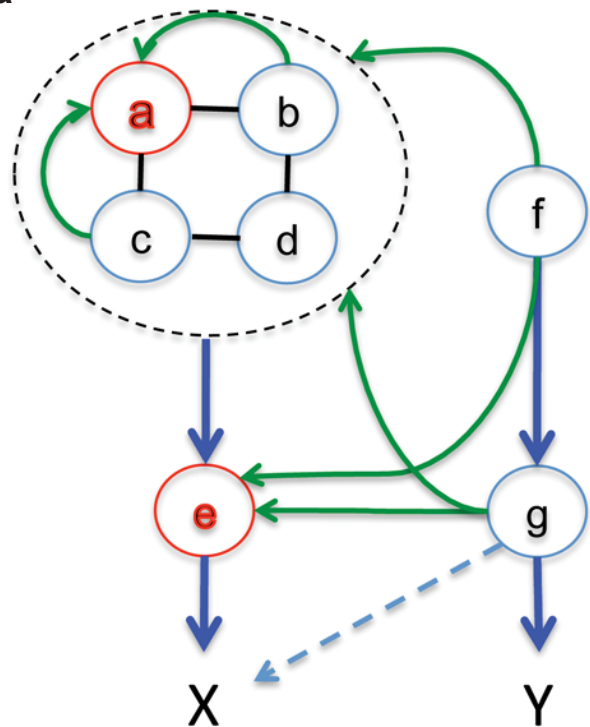
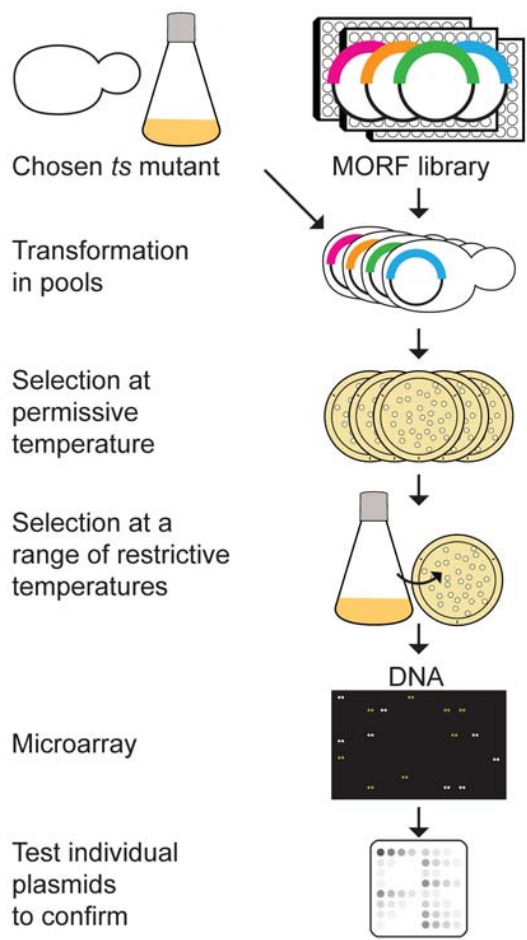
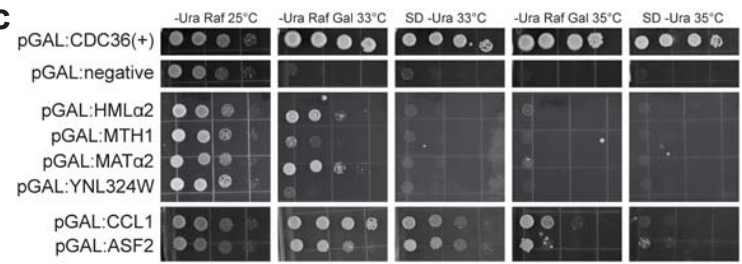
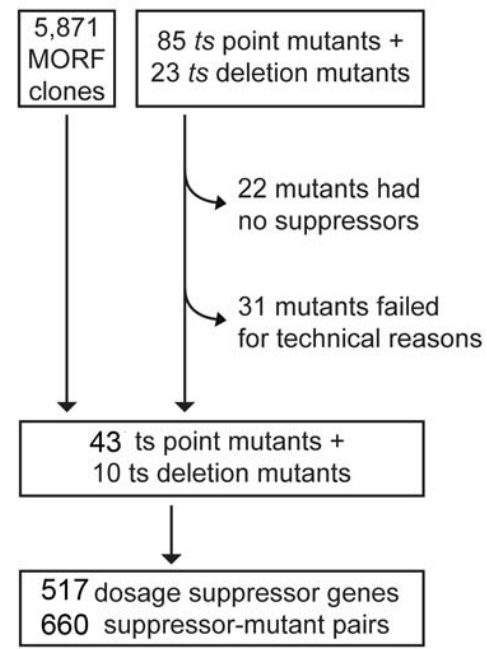

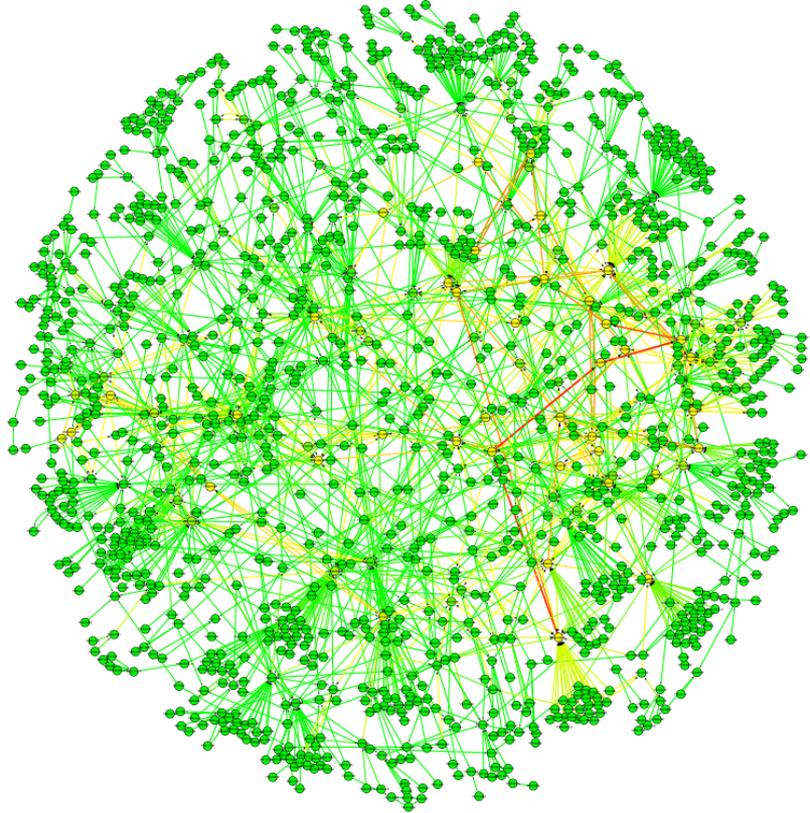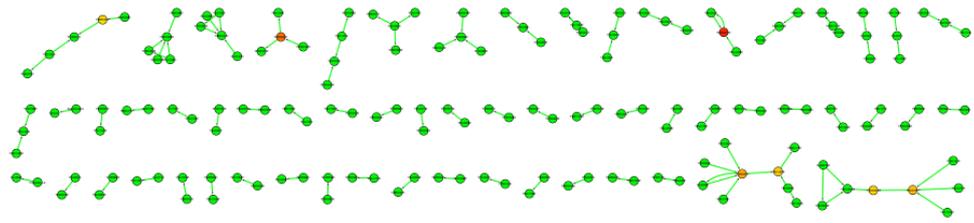

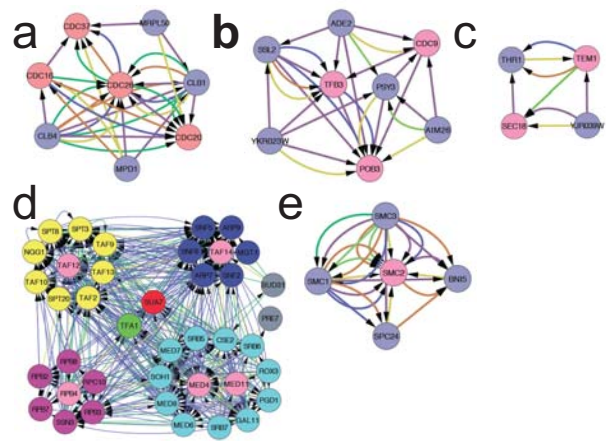

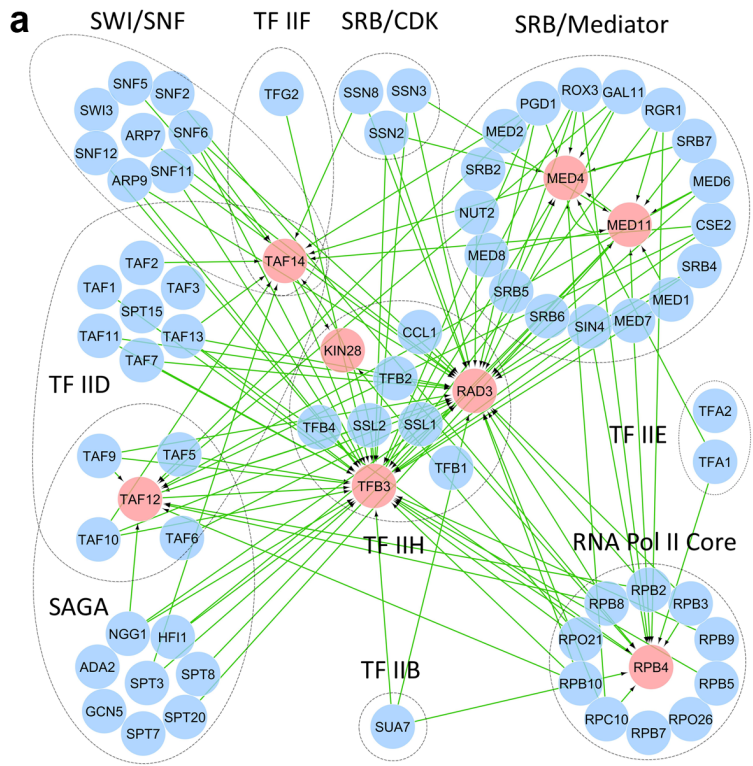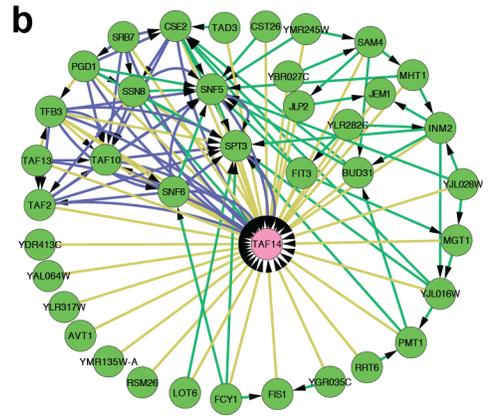

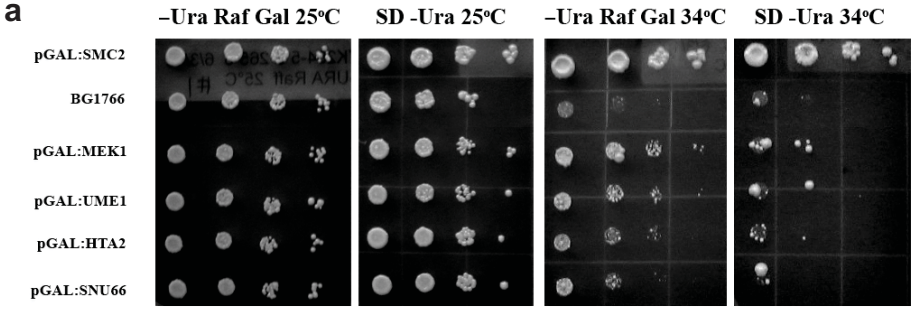
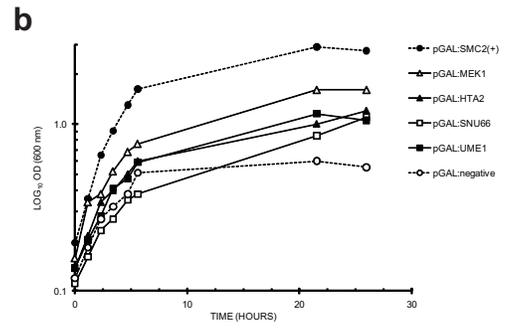
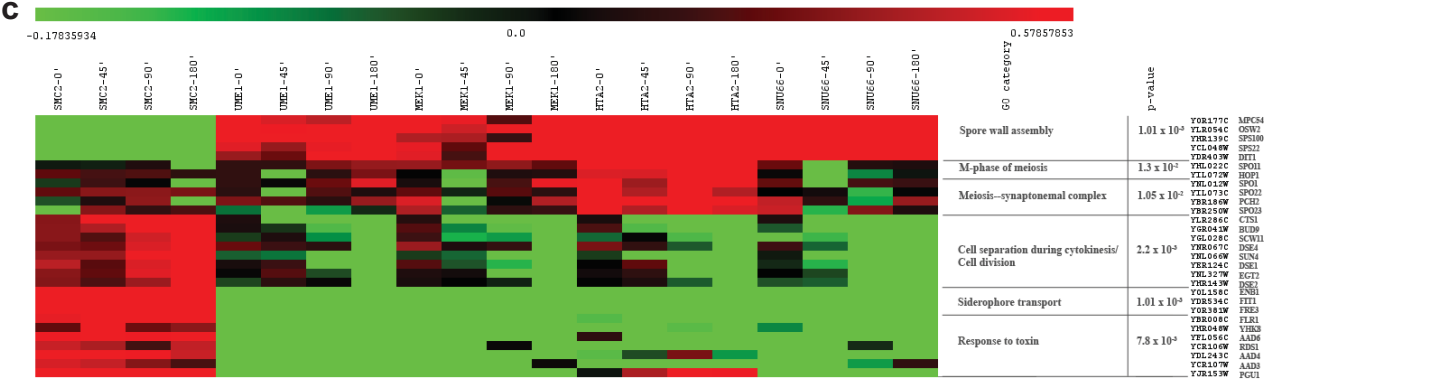
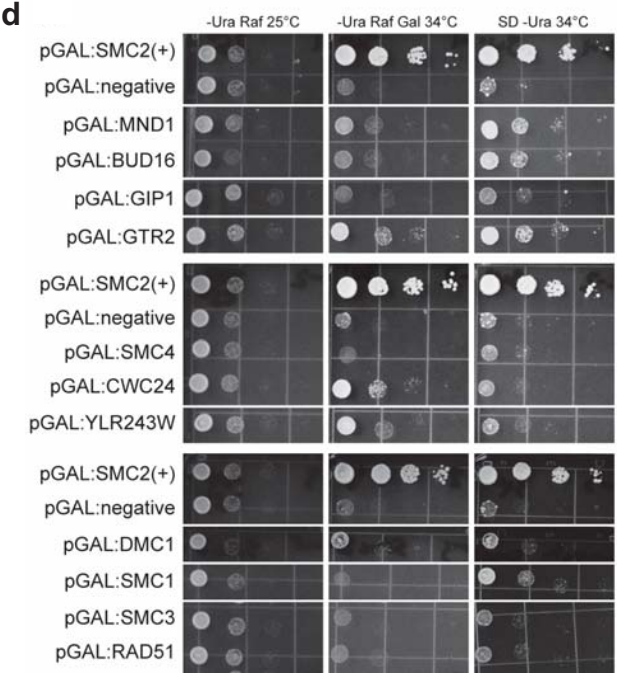
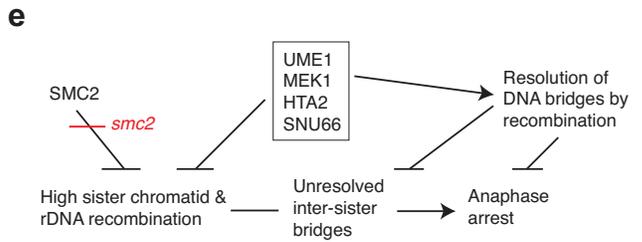

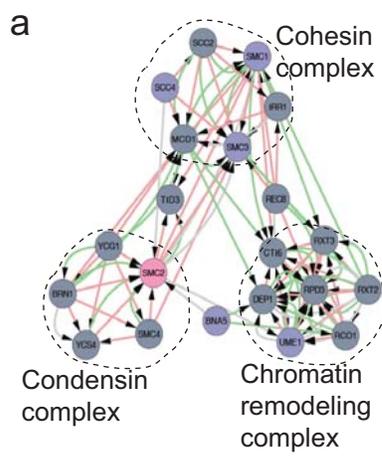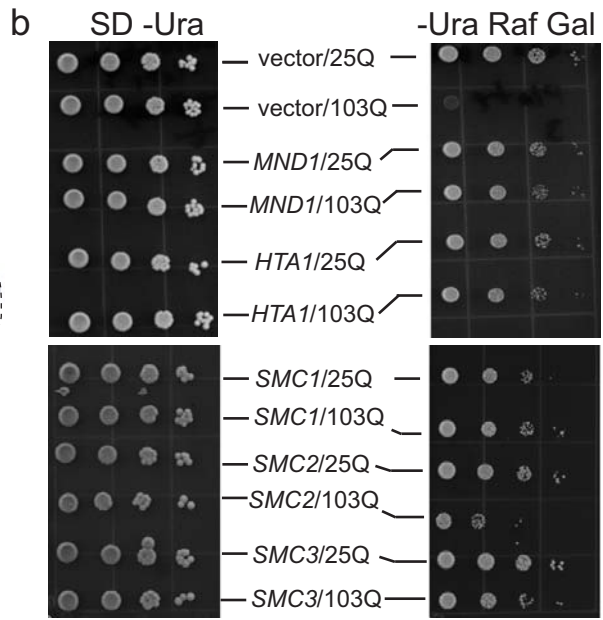